\documentclass [12pt,english]{article}
\usepackage{amssymb}
\usepackage{amsmath}
\usepackage{amsfonts}
\usepackage[english]{babel}

\usepackage[T2A]{fontenc}
\usepackage[cp1251]{inputenc}
\usepackage{fullpage}
\usepackage{bm}
\usepackage{graphicx}
\numberwithin{equation}{section}
\newcounter{Alpha}

\begin{document}

\title{\LARGE 
Quantum Phase Transitions in 2D--dimensional paramagnetic half--filled
Hubbard model\\}

\author{N. I. Chashchin\thanks{E--mail: nik.iv.chaschin@mail.ru} \\ \textit{Ural State Forestry University}\\
\textit {Ekaterinburg, Sibirskii trakt 37, 620100 Russia}}
\date{}

\maketitle

\begin{abstract}

We investigate the quantum phase transitions  in strongly correlated 
electronic systems at $T=0^0K$ by the example of the 2D Hubbard model. The model for numerical calculations were formalized in terms of the integral equations previously obtained for the half--filled paramagnetic Hubbard model by means of the variational derivatives technique and  subsequent Legendre transformation.
For principal results we submit in series the momentum distribution functions $n(k)$ which display distinct attributes for the incipient regimes: the pure doublon phase, the metal  phase, Weyl semimetal phase, and Weyl excitonic insulator phase at the sequential increasing the onsite Coulomb interaction $U$.

\end{abstract}

{\bf Keywords:} Hubbard model,  quantum phase transition, electronic spectrum, 
doubly occupied sites, momentum distribution, Green's functions, exciton, generating functional, Legendre transformation. 
\\
{}
\\
DOI:10.1134/S0031918X1607036\\
Pacs numbers 31.15.xt; 31.15.xp; 71.10.Fd

\section{\!\!\!\!\!\!. Introduction}

In past decades the phase transitions driven by fermionic correlations ---  
quantum phase transitions (QPTs) have attracted the interest of many 
researchers in the field of strongly correlated electronic systems \cite{{Sachdev},{Vojta}, 
{Irkhin},{Sachdev_2}}. These transitions as opposed to classical phase transitions
are taken place at zero temperature by varying a non-thermal control
parameter, but some parameter of the Hamiltonian. Thus QPTs occur as a result of competing ground state phases and their critical behaviour must be definitely treated quantum mechanically. 

In this article we study the quantum phase transitions by the example of the paramagnetic 2D repulsive Hubbard model at half filling, where the electronic Coulomb interection U is a control parameter.

Interplay of band theory and strong electron correlation has been a basic phenomenon
for understanding physics of many--body models. Electron repulsion due to local Coulomb interactions tends to localize electrons, on the other hand, kinetic effects are
favourable to electron itineracy. 

A distinguishing characteristic of strongly correlated phases in materials is that the behavior of their constituent electronic excitations can no longer be described as a non-interacting renormalized Fermi gas. The energetic competition between them leads to different quantum phases of comparable magnitude. The apparent strong nonlinear nature of the problem requires  nonperturbative treatment of the model.

The single band Hubbard model (HM) with one electron per site (halfilled) is the simplest model and  archetypal example for investigation of correlated electrons. 
Originally, the Hubbard model was introduced  
for describing electronic correlations in narrow energy bands of transition metals but subsequently it turned out to be a very simple pattern and benchmark for understanding of the unusual electronic properties of strongly correlated systems \cite{Hubbard1963}.

The initial band of noninteracting electrons is splitted into two subbands in a correlated state: the lower -- valence band and the upper -- conductivity band. The electronic spectrum of a crystal can be organized in the form of energy bands as a function of the crystal momentum $k$, which is
guaranteed to be a good quantum number due to the translational symmetry of the lattice.
When an electron is excited from the valence into the conductivity band it leaves behind a
vacancy -- a hole. Electrons are concentrated near minimums of the
conductivity band and holes near maximums of the valence band. Nearby these extremums
band energy spectra in simple form is $E(k)\approx\hbar^2 k^2/2 m^*$ , where $m^*$ is an effective mass of the electrons or the holes correspondingly. The vacant place (hole) in the filled valence band has the negative mass in definition, though it is convenient to consider this hole as a real particle with the positive mass and charge, and it is possible treatng the electrons and holes as their own statistical mechanical systems.

A great number of remarkable dynamical phenomena have been found out, including a correlation--driven Mott transition \cite{{Mott},{Schafer}}, and mass enhancement 
\cite{Dordevic}(heavy fermions) at the Fermi level; as well as non-Fermi-liquid properties, a pseudogap in the electronic spectrum, and so on. It is indicative that angle--resolved photoemission studies (ARPES) of materials report unusual quasiparticle properties at all carrier concentrations \cite{Maier}. 

The Hubbard model, where Coulomb interactions between particles are assumed to be local, has been commonly used to study Mott localization and the related correlation-induced effects. Understanding these effects  
and the associated with it localization of charged fermions --- electrons and holes remains to be a problem of interest in condensed matter physics.

The local double occupancy (doublon) is a correlation function of a pair of electrons  
$n^e_{1\sigma}$ 
with antiparallel spins $\langle n^e_{1\uparrow}n^e_{1\downarrow}\rangle$ on one site.
The double occupancy plays an important role in the study of correlated systems. It critically influences on the Mott metal (semimetal)--insulator transition and the local moment formation. In optical lattice experiments, the double occupancy gives information about a physical phase of the system \cite{{Tudor},{Erik}}. 
The local Coulomb repulsion U and the kinetic mobility of electrons disfavors doubly-occupied sites and hence decreases the function.
 
It is appropriate to assume that there also has to be a local bound state of an electron in the conductivity band and a hole in the valence band are attracted to each other by the Coulomb force and forms a neutral quasiparticle that exists in various electronic phases of the model and which is named as an exciton. The condensation of excitons in a macroscopic quantum state has been proposed soon after the success of BCS theory of superconductivity \cite{Keldysh} owing to the similarities between the Cooper pairs created by the binding of two electrons, and the excitons -- bound states formed by an electron and a hole.
 
In systems where electron and hole states are mixed at the Fermi level --- that is typical for the semimetal phase --- the Coulomb interaction between the oppositely charged electrons $n^e_{1\sigma}$ and holes 
$n^h_{1\sigma}=1-n^e_{1\sigma}$ leads to the formation of charge--neutral electron--hole pairs $\langle n^e_{1\uparrow}n^h_{1\downarrow}\rangle$  
which produce a set of hydrogen-like subjects --- the localized on a site Frenkel-type excitons (which are analogous to hydrogen atoms) 
\cite{{Plakida},{Zenker},{Mazziotti}}.

The local Coulomb repulsion increases the electron--hole double occupancy in contrast to the electron--electron one. 
The correlation effects caused by these bosonic--type single quantum states are of an interest in the study of electronic systems. 

The momentum distribution function $n(k)$, or Fermi function provides the probability of an average occupation number of the quantum state with lattice momentum k. The relevance of this function is related to the fact that its analitical properties are generally considered for discerning among different classes of electronic liquids \cite{Kotliar,Brech} and it also 
characterizes other electronic correlations. 
The maximum energy of an electron at  $T=0^0K$ is known as Fermi energy level $E_F$. It is possible from view of the function to specify a particular quantum  phase of the electronic  system.
 
\section{\!\!\!\!\!\!. Model and Method} 

In the simplest form, the Hamiltonian of the Hubbard model is written as

\begin{equation}
\mathcal{H}= -t\sum\limits_{\langle i,j\rangle\sigma}c_{i\sigma}^{\dag}c_{j\sigma}+\sum\limits_{i\sigma}
\varepsilon_\sigma n_{i\sigma}
+U\sum\limits_{i}n_{i\uparrow}n_{j\downarrow}\,,
\label{Pm:Hub_ham}
\end{equation} 
where $U$ is the parameter of the Coulomb interaction at a site;  
$c_{i\sigma} (c_{i\sigma}^\dag)$ are  the Fermi operators describe 
the annihilation (generation) of electrons with spins up and down 
$\sigma=\uparrow,\downarrow$; 
 $n_{i\sigma}$  indicates the operators of the number of particles; $t$ is the parameter 
of hopping of s--electrons from site to site; in the designation
$\langle i,j\rangle$ sites are nearest; $\varepsilon_\sigma=-\displaystyle{\sigma\frac{h}{2}-\mu}$,
where $h=g\mu_BH$ and $g$ is the electronic $g-$factor, $\mu_B$ is the Bohr magneton; 
$H$ is the external magnetic field, and $\mu$ is the chemical potential.

In our previous works \cite{chaschin2011_2,chaschin2012_3,chaschin2016_4,chaschin2011_1,izyumov_Manc2005,
chaschin2020}
we had obtained by means of generating functional  method and the subsequent Legendre transformation the mathematical model for HM  represented by two connected sets of the singular  integral equations \cite{chaschin2017} for two types of Green functions. 
The electron propagator 
\begin{equation}
\displaystyle G({\bf k}, i\omega_n)=\frac{1}{i\omega_n-\varepsilon_{\bf k}-\Sigma({\bf k},i\omega_n)}\,,
\label{N_el}
\end{equation} 
where $\varepsilon_{\bf k}$ is the electronic free spectrum, 
$\omega_n=(2n+1)\, \pi T$ ($n=0,\pm 1,\pm 2,\dots$) are fermionic Matsubara frequencies, and 
$\Sigma({\bf k},i\omega_)$ is the electronic self energy;  
and the bosonic propagator 

\begin{equation}
\displaystyle Q({\bf q},i\Omega_\nu)=-\frac{1}{1+\frac{U}{2}\Pi(q,i\Omega_\nu)}\,,
\label{Q_bos}
\end{equation}
where $\Omega_\nu=2\, \nu \,\pi T$ ($\nu=0,\pm 1,\pm 2,\dots$) are bosonic Matsubara frequencies, and 
$\Pi({\bf q},i\Omega_\nu)$ is the bosonic self energy.
 
The lattice dimension D enters into the equations as some external parameter not necessarily as an integer number and determines the screened interaction constant
 
\[U_{eff}\sim\frac{U}{2^{D-2}\pi^{(D-1)/2}\Gamma{[(D-1)/2}]}\]\, 
which specifies the electronic correlations in the system.

\begin{figure}[h]
\begin{center}
\includegraphics[width=0.6\textwidth]{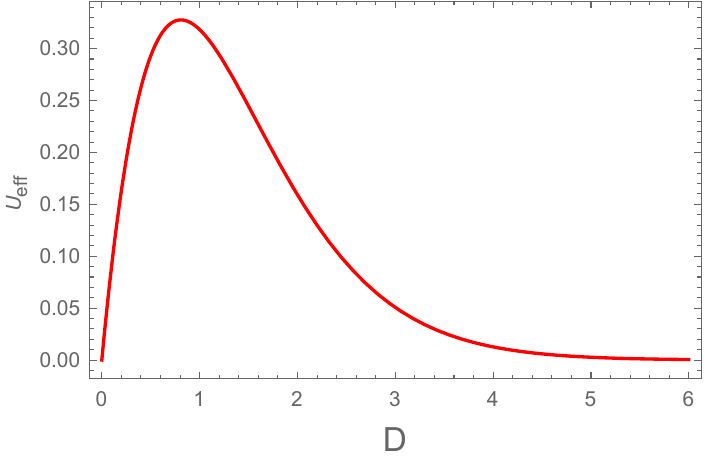}
\caption{The effective Coulomb interaction constant $U_{eff}$ as a function of $D$. 
For $D>4$ $U_{eff}\sim 0$,  so $D=4$ can be regarded as some threshold value for the correlation effects.} 
\label{Pm:Cd}
\end{center}
\end{figure}

Fig.\ref{Pm:Cd} shows substantial D--dependence of
$U_{eff}$. 
 
In that way we get resources to investigate directly the influence of lattice dimensiality on some correlation phenomena of the model. 


\section{\!\!\!\!\!\!. Results and Discussion}
\addtocounter{Alpha}{1}
{\bf\large{\Alph{Alpha}. Electronic double occupancy (doublons) and localized exciton states}}\\ 
\setcounter{Alpha}{1}
\addtocounter{Alpha}{1}

To get the expression for the average number of doublons
write down in the beginning the two--particle Green function in the explicit form 
\begin{equation}
\displaystyle \langle \hat T\,n_{1\uparrow}n_{1\downarrow}\rangle = 
\displaystyle \langle (n_{1\uparrow}-\frac{1}{2})(n_{1\downarrow}-\frac{1}{2})\rangle = 
\langle n_{1\uparrow}n_{1\downarrow}\rangle -\frac{\langle n_1\rangle}{2}+\frac{1}{4}\,,
\label{A_Dbl}
\end{equation}
as in our symmetrical representation 
$\displaystyle\hat T n_{1\sigma}=n_{1\sigma}-\frac{1}{2}$. 

From the other hand using the variational derivatives technique \cite{chaschin2011_1}, 

\begin{equation}
\displaystyle \langle \hat T\,n_{1\uparrow}n_{1\downarrow}\rangle = 
\frac{1}{4}\left[\frac{\delta^2\Phi}{\delta \rho(11)\,{\delta\rho(11)}}-
\frac{\delta^2\Phi}{\delta \eta(11)\,{\delta\eta(11)}}\right]+ 
\frac{1}{4}\left[\left(\frac{\delta\Phi}{\delta \rho(11)}\right)^2 - 
\left(\frac{\delta\Phi}{\delta \eta(11)}\right)^2\right]\,,
\label{A_Dbl_GF}   
\end{equation}

where $\Phi=\ln Z$ is the generated functional of the connected Green functions; $\displaystyle\frac{\delta\Phi}{\delta \rho(11)}=\langle n_1\rangle-1$, 
$\displaystyle\frac{\delta\Phi}{\delta \eta(11)}=\langle n_{1\uparrow}-n_{1\downarrow}\rangle$;
 $1\equiv (i,\tau)$, i--a crystal site, 
$\tau$--an imaginary time; $\eta(12)=h\,\delta_{12}$ is the external magnetic field; 
$M_1=\langle n_{1\uparrow}-n_{1\downarrow}\rangle$ --- the local magnetic moment.

As we see from (\ref{A_Dbl_GF}), determinal factors for the propagator of double occupied sites are two--particle Green functions:   
\begin{equation}
\displaystyle \langle \hat T\, M_1^2\rangle=\frac{\delta^2\Phi}{\delta \eta(11)\,{\delta\eta(11)}},\;\;
\displaystyle \langle \hat T\, n_1^2\rangle=\frac{\delta^2\Phi}{\delta \rho(11)\,{\delta\rho(11)}}\,.
\label{DPM:XM}   
\end{equation}

Pauli exclusion principle in the functional form \cite{chaschin2011_1} is  
\[
\displaystyle\frac{\delta^2\Phi}{\delta \rho(11)\,{\delta\rho(11)}}+
\frac{\delta^2\Phi}{\delta \eta(11)\,{\delta\eta(11)}}+
\left(\frac{\delta\Phi}{\delta \rho(11)}\right)^2+
\left(\frac{\delta\Phi}{\delta \eta(11)}\right)^2 = 0,
\]  

which allow to write (\ref{A_Dbl_GF}) in the form 
\begin{equation}
\displaystyle\langle \hat T\,n_{1\uparrow}n_{1\downarrow}\rangle =
-\frac{1}{2}\,\frac{\delta^2\Phi}{\delta \eta(11)\,{\delta\eta(11)}}-
\frac{1}{2}(\langle M_1\rangle)^2\,.
\label{A_Dbl_2}
\end{equation}
Comparing two expressions (\ref{A_Dbl}, \ref{A_Dbl_2}) we get general formula for the number of double--occupied sites 

\begin{equation}
\displaystyle\langle n_{1\uparrow}n_{1\downarrow}\rangle = \frac{\langle n_1\rangle}{2}
-\frac{1}{4}-\frac{\langle M_1\rangle^2}{2}- 
\frac{\langle\hat T\, M^2_1\rangle}{2}\,.
\label{A_Dbl_M}
\end{equation}

The mathematical expression for $\langle\hat T\, M^2\rangle$ was earlier derived in 
\cite{chaschin2016_4}:

\begin{equation}
\displaystyle \langle\hat T\, M^2\rangle =
-\frac{1}{\pi U}\sum\limits_{k}\int_{-\infty}^\infty 
\tanh\left(\frac{\omega}{2 T}\right)\left(\omega-\varepsilon(k)\right)\,\Im G(k,\omega)\,d\omega\,
\label{Pm:MM}
\end{equation}

Finally for paramagnetic $\langle M\rangle=0$ and half--filled $\langle n\rangle=1$ solution the average number of double--occupied sites (\ref{A_Dbl_M}) is  
\begin{equation}
\displaystyle \langle n^e_{\uparrow}n^e_{\downarrow}\rangle=\frac{1}{4}-\frac{\langle\hat T\, M^2\rangle}{2} 
\equiv\mathcal{X}(U)\,,   
\label{PM:Dbl}   
\end{equation}
where we denote now $n^e_{\sigma}$ as the operator for electrons. 

\begin{figure}[h]
\begin{center}
\includegraphics[width=0.6\textwidth]{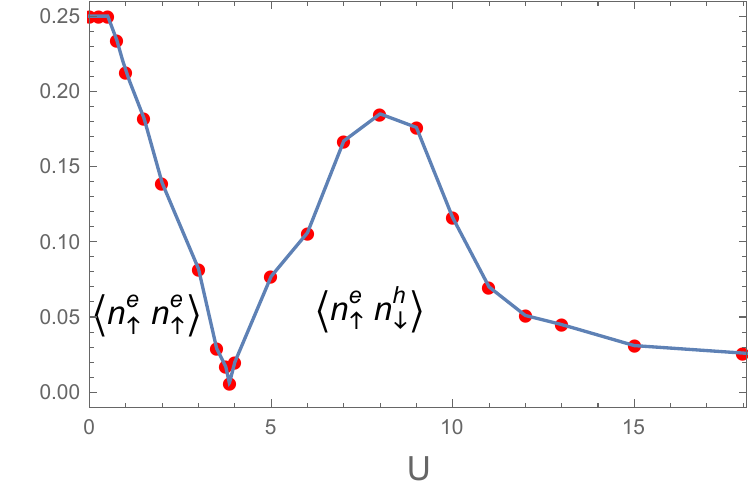}
\caption{The function of an average 
number of sites with the double occupancy by electrons 
Dbl=$\langle n^e_{1\uparrow}n^e_{1\downarrow}\rangle$, and an average number of sites which occupied by the localized excitons 
Exc=$\langle n^e_{1\uparrow}n^h_{1\downarrow}\rangle$. At $U\approx 4$ there takes place a sharp change-over from doublons to excitons.} 
\label{Pm:Dbl}
\end{center}
\end{figure} 

Put the next graphic designations:
\unitlength=1mm \,
\begin{picture}(4,6)
\put(2,1){\circle{3}}
\end{picture} is just a vacant place uncorrelated with other fermions, and unbound to a host--site; 
\begin{picture}(4,6)
\put(2,1){\circle{3}}
\put(.5,1){\line(1,0){3}}
\put(2,2.5){\vector(0,1){3}}
\end{picture} is a negative charged electron; 
\begin{picture}(4,6)
\put(2,1){\circle{3}}
\put(2,2.5){\vector(0,1){3}}
\put(.5,1){\line(1,0){3}}
\put(2,-.5){\line(0,1){3}}
\end{picture} is a positive charged hole.
 
In that case all electronic configurations on a single site can be represented in one of the particular form 
\unitlength=1mm \,
\begin{picture}(57,10)
\put(0,1){\circle{3}}
\put(3.2,1){\circle{3}}
\put(6,0){,}
\put(10,1){\circle{3}}
\put(13,1){\circle{3}}
\put(10,2.5){\vector(0,1){3}}
\put(13,-.5){\vector(0,-1){3}}
\put(8.5,1){\line(1,0){3}}
\put(11.5,1){\line(1,0){3}}
\put(16,0){,}
\put(20,1){\circle{3}}
\put(23.2,1){\circle{3}}
\put(20,2.5){\vector(0,1){3}}
\put(18.5,1){\line(1,0){3}}
\put(26,0){,}
\put(30,1){\circle{3}}
\put(33.2,1){\circle{3}}
\put(33.2,-.5){\vector(0,-1){3}}
\put(31.5,1){\line(1,0){3}}
\put(33.2,-.5){\line(0,1){3}}
\put(36,0){,}
\put(40,1){\circle{3}}
\put(43.2,1){\circle{3}}
\put(39.8,2.5){\vector(0,1){3}}
\put(43.2,-.5){\vector(0,-1){3}}
\put(38.5,1){\line(1,0){3}}
\put(41.5,1){\line(1,0){3}}
\put(43.2,-.5){\line(0,1){3}}
\put(46,0){,}
\put(50,1){\circle{3}}
\put(53.2,1){\circle{3}}
\put(50,2.5){\vector(0,1){3}}
\put(53.2,-.5){\vector(0,-1){3}}
\put(48.5,1){\line(1,0){3}}
\put(51.5,1){\line(1,0){3}}
\put(50,-.5){\line(0,1){3}}
\put(53.2,-.5){\line(0,1){3}}
\put(56,0){.}
\end{picture}\,\,

Using the operator expression for a hole's number $n^h_{\sigma}$= $1-n^e_{\sigma} $ we can write the   
obvious identity for the average number of electrons with a certain spin direction in the system as  
$\displaystyle 
\langle n^e_{\sigma}\rangle=\langle n^e_{\sigma}n^e_{\bar\sigma}\rangle +\langle n^e_{\sigma}n^h_{\bar\sigma}\rangle$; from which we get   

\begin{equation}
\begin{array}{l}
\displaystyle\langle n^e_{\sigma}n^e_{\bar\sigma}\rangle = \mathcal{X}(U)
\\ 
\displaystyle\langle n^e_{\sigma}n^h_{\bar\sigma}\rangle - \langle n^e_{\sigma}\rangle = -\mathcal{X}(U)\,.
\end{array}
\label{Dbl_Exc}
\end{equation}

A numerical value of $\langle n^e_{\uparrow}n^h_{\downarrow}\rangle$ can be expressed in the graphic view as  
$\displaystyle\langle n^e_{\uparrow}n^h_{\downarrow}\rangle$= 
$\langle\,
\begin{picture}(7,7)
\put(2,1){\circle{3}}
\put(5.2,1){\circle{3}}
\put(2,2.5){\vector(0,1){3}}
\put(5,-.5){\vector(0,-1){3}}
\put(.5,1.2){\line(1,0){3}}
\put(3.5,1.2){\line(1,0){3}}
\put(5,-.5){\line(0,1){3}}
\end{picture}\,\rangle$ + 
$\langle\,
\begin{picture}(7,7)
\put(2,1){\circle{3}}
\put(5.2,1){\circle{3}}
\put(2,2.5){\vector(0,1){3}}
\put(.5,1.2){\line(1,0){3}}
\end{picture} \,
\rangle.$
At that $\langle\,
\begin{picture}(7,7)
\put(2,1){\circle{3}}
\put(5.2,1){\circle{3}}
\put(2,2.5){\vector(0,1){3}}
\put(.5,1.2){\line(1,0){3}}
\end{picture} \,
\rangle$ = $\langle n^e_\uparrow\rangle$, and 
$\langle\,
\begin{picture}(7,7)
\put(2,1){\circle{3}}
\put(5.2,1){\circle{3}}
\put(2,2.5){\vector(0,1){3}}
\put(5,-.5){\vector(0,-1){3}}
\put(.5,1.2){\line(1,0){3}}
\put(3.5,1.2){\line(1,0){3}}
\put(5,-.5){\line(0,1){3}}
\end{picture}\,\rangle$ = $\langle\mathcal N^\text{exc}_\uparrow\rangle$ -- an excitonic pair for the single 
($\uparrow$)  spin direction.

From the second equation of (\ref{Dbl_Exc}) we finally find the average number of localized excitons for both spin directions as 
\begin{equation}
\displaystyle\langle\mathcal N^\text{exc}\rangle = - 2\,\mathcal{X}(U)\,.
\label{N_exc}
\end{equation} 
Positive or negative value of the parameter $\mathcal{X}(U)$ for different U specifies the electronic state of the system, its quantum phase.  The interaction drives the bands from a semimetallic configuration to an excitonic insulating one. The excitonic nature of the new state is revealed itself in the deformed and mixed valence and conductivity subbands, which are indicating the hybridization of the original band and the correlated coexistence of electrons with holes (see Fig.\ref{Pm:Dbl}).


\addtocounter{Alpha}{0}
{\bf\large{\Alph{Alpha}. 
Momentum distribution functions and quantum phase transitions}}\\ 
\setcounter{Alpha}{0}
\addtocounter{Alpha}{0}

One of the hallmark parameteres which reveals the correlation in the electronic system presents the momentum distribution function $n(k)$, or Fermi function. 
It provides the probability of an average occupation number of electronic states  of the momentum $k$ at some temperature \cite{Kotliar,Brech}. The maximum energy of an electron at  $T=0^0K$ is known as Fermi energy level $E_F$. 

The Fermi function per one spin direction for the ground state can be expressed in terms of the Matsubara Green function as 
\begin{equation}
\displaystyle n(k)=-\frac{1}{2\pi}
\displaystyle\int_{-\infty}^\infty \tanh\left(\frac{\omega}{2 T}\right)\,\Im G(k,\omega)\, 
d\omega\,. 
\label{DPM:nk}
\end{equation}
The relevance of the function (\ref{DPM:nk}) refers to its analitical properties \cite{Meixner}. 
The four panels 
Fig.\ref{Pm:nk0}, Fig.\ref{Pm:nk1}, Fig.\ref{Pm:nk2}, Fig.\ref{Pm:nk3} exhibit  
the momentum distribution graphixs  as functions of $(k-k_F)/2k_F$, the red line corresponds 
to free electrons.
Every of four panels corresponds to a particular physical state of the electronic system, i.e. we are dealing with some sort of quantum phase transitions by variation of U as a control parameter. 
Their forms and configurations are varying substantially for different values of the constant U. Apparently condensate fluctuations of the bosonic type
influence on the behaviour of fermions in the system and conversely  fermions
influence strongly on low energy behaviour of the bosonic parameters.

\section{\!\!\!\!\!\!. Summary and Discussion}

The basic purpose of this paper was an investigation of physical properties of the half--filled paramagnetic 2D--Hubbard model at zero temperature when electronic Coulomb repulsion U on a site is changing.

The model was formalized in terms of  the integral equations, which were obtained previously \cite{chaschin2017} by means of the variational derivatives technique and  subsequent  Legendre transformation. 

\begin{figure}[h]
\begin{center}
\includegraphics[width=.6\textwidth]{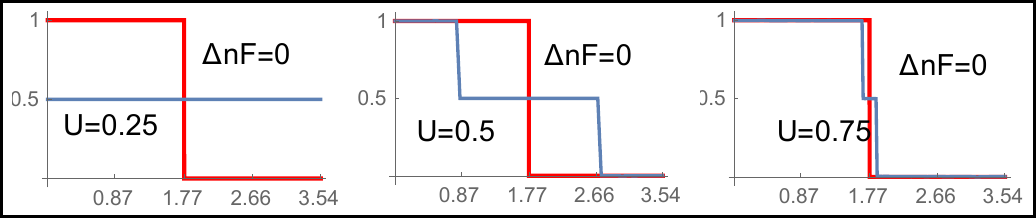}
\caption{This panel  
exhibits the momentum distributions as function of $(k-k_F)/2k_F$ for small values of 
$U$=0.25, 0.5, 0,75. $\Delta F$=$n(k_F-0)-n(k_F+0)$ 
is a jump at the Fermi level, the red line corresponds 
to free electrons. The k--region of bound electron--electron condenced fluctuations gradually reduces to zero keeping behind residual doublons (Dbl) and almost free electrons. Peculiar view of the graphixs can be explained by strong influence of  the bosonic fluctuations on the electronic behaviour at low Coulomb interaction.}    
\label{Pm:nk0}
\end{center}
\end{figure} 

\begin{figure}[h]
\begin{center}
\includegraphics[width=.8\textwidth]{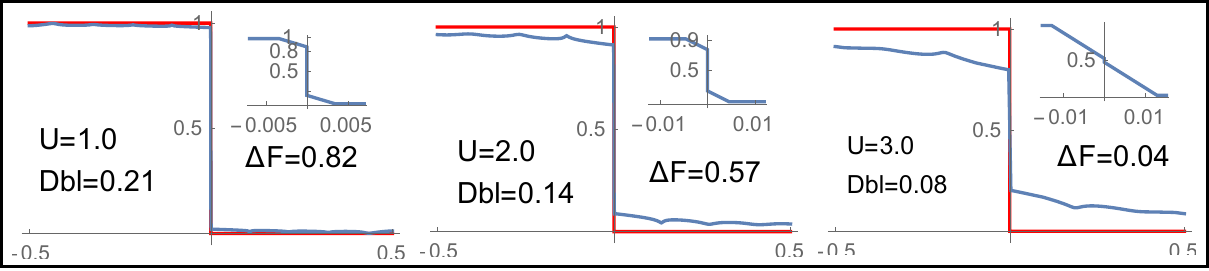}
\caption{The panel exhibits the Fermi functions for U=1.0, 2.0, 3.0. At U$\simeq$1
the boson electron--electron fluctuations vanish entirely from the system that is the doublons 
(Dbl) are disintegrated into the system of correlated electrons.  
$\Delta F$=$n(k_F-0)-n(k_F+0)$ has a positive rapidly decreasing jump of $n(k)$ at the Fermi level, that corresponds to the metal phase.}
\label{Pm:nk1}
\end{center}
\end{figure} 

\begin{figure}[h]
\begin{center}
\includegraphics[width=.8\textwidth]{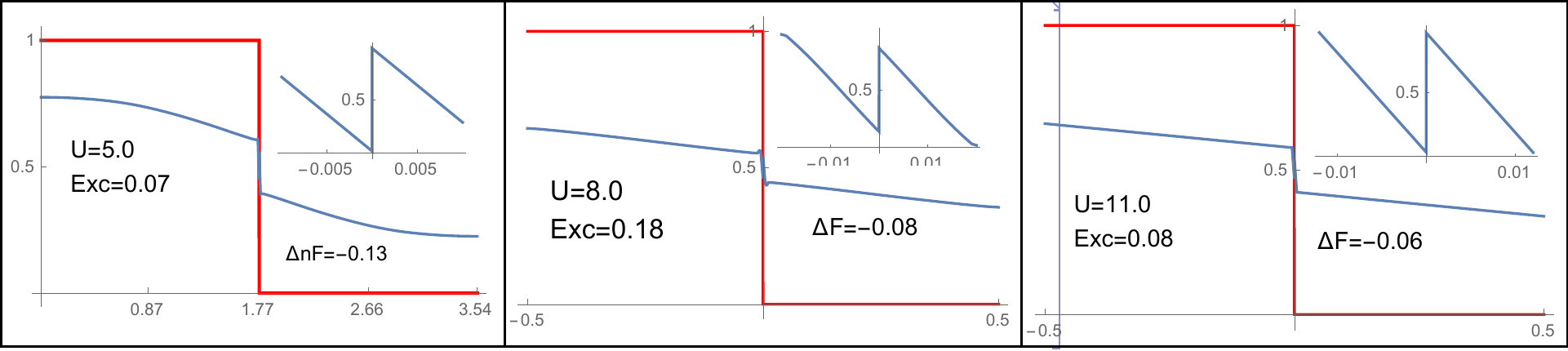}
\caption{The panel exhibits the Fermi functions for U=5, 8, 11. Here we observe small overlap between the bottom of the conduction band and the top of the valence band. The jump 
$\Delta F$  is negative at the Fermi level  with sharply varying values of excitons localized on sites (Exc).  The band inversion regime is inherent in so called Weyl semimetals.}
\label{Pm:nk2}
\end{center}
\end{figure} 

\begin{figure}[h]
\begin{center}
\includegraphics[width=.8\textwidth]{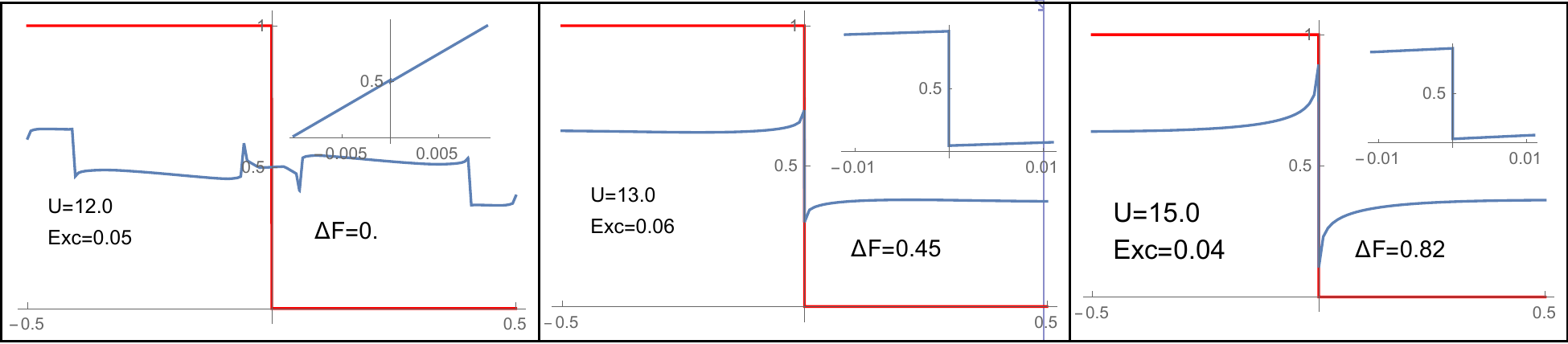}
\caption{The panel exhibits the Fermi functions for quite big U = 12, 13, 15. The sharp rise of the function in the valence band and the sharp fall in the conduction band produces an energy gap. The gap $\Delta F$ has a positve and rapidly increasng jump of $n(k)$ at the Fermi level. The first graphicx corresponds to the onset of the quantum transition from Weyl semimetal to Weyl insulator.}  
\label{Pm:nk3}
\end{center}
\end{figure} 

For that purpose it was numericaly calculated the one-- and two--particle electronic Green functions and momentum distribution functions $n(k)$. It had turned out that these Fermi functions are essentially differed in features at change of U. We had observed the four  regimes, where behaviour of the functions $n(k)$ were fundamentally distinguished. In such a way here appeared the new insipient quantum phases, crossovers between  them  are quite smooth and one can conclude that  
we were dealing with subsequent quantum phase transitions and Coulomb interaction U is the control parameter of the transitions 
{Fig \ref{Pm:nk0}, Fig \ref{Pm:nk1}, Fig \ref{Pm:nk2},Fig \ref{Pm:nk3}}.

In Fig \ref{Pm:Dbl} it is shown the function of an average 
number of sites with the double occupancy by electrons 
$\langle n^e_{1\uparrow}n^e_{1\downarrow}\rangle$, and an average number of sites which occupied by the localized excitons 
$\langle n^e_{1\uparrow}n^h_{1\downarrow}\rangle$. At U$\simeq$4 there takes place a sharp change-over from doublons to excitons. 

So we have four different regions in the system and the determinal factor for them is a behaviour of particles near the Fermi level. 

\begin {enumerate}
 
\item 0<U<1

In this region simultaneously exists uncorrelated electrons and paired up ones into the boson-type  double occupation excitations. The physical properties of the system for very small U<0.2 is determined exclusivly by localized boson excitations--doublons. On the increasing of U there occurs a gradual displacement of the decomposing doublons by weakly correlated electrons (see Fig \ref{Pm:nk0}). At U$\simeq$1 we get into the phase with nearly free unmixed electrons and holes.

\item 1<U<3

Fig \ref{Pm:nk1} corresponds to a regime, where at the increasing of U a positive jump $\Delta F$ of the momentum distribution function $n(k)$ at the Fermi level is decreasing to zero at U$\simeq$3. At this range we are dealing with different metallic phases which  
gradually get into the next quantum phase.

\item 3<U<11

Fig \ref{Pm:nk2} corresponds to a regime where the Fermi function $n(k)$ has a negative jump at the Fermi level. Obviously the negative jump is conditioned by 
a small overlap between the bottom of the conduction band and the top of the valence band, i.e. the subbands are mixed. This phase with the inversion jump is inherent in so called Weyl semimetal. The further rising of U drives the semimetal to undergo a quantum phase transition into an excitonic insulator \cite{Pan}.
 
\item U>11

In the last range (see Fig \ref{Pm:nk3}) we observe an energy gap between the valence and conduction bands of the system. The sharp rise of the function in the valence band and the sharp fall in the conduction band produces the uprising energy gap in the electronic spectrum. 
The gap $\Delta F$ has a positve and rapidly increasng jump of $n(k)$ at the Fermi level. The phase corresponds to so called Weyl insulator.

\end{enumerate}

Thus we subsequently  have described a number of
specific quantum phase transitions in the considered Hubbard model, drawing the special  attention to the important role of low-energy fermionic excitations.


\begin{thebibliography}{21}

\bibitem{Hubbard1963} {J. Hubbard}, J.Proc.Roy.Soc.A {\bf{276}}, 238 (1963) 

\bibitem{Sachdev}{S. Sachdev and B. Keimer}, Physics Today 64, 29 (2011). 

\bibitem{Vojta}{M.Vojta},  arXiv:0309604v2 [cond-mat.str-el], (2003)

\bibitem{Irkhin}{V.Yu.Irkhin and Yu.N.Skryabin}, arXiv:1909.06248v1 [cond-mat.str-el] (2019)

\bibitem{Sachdev_2}{S. Sachdev},  arXiv:2407.15919v1 [cond-mat.str-el], (2024)

\bibitem{Mott}{N.F.Mott}, Metal-Insulator Transitions Taylor and Francis London (1990).

\bibitem{Schafer}{T. Sch\"{a}fer et al.}, Phys. Rev. B. 91, 125109 (2015).

\bibitem{Dordevic}{S.V. Dordevic }, arXiv:2407.17290v1 [cond-mat.str-el] , (2024)

\bibitem{Maier}{Th.A. Maier,�Th. Pruschke, and�M. Jarrell},
Phys. Rev. B 66, 075102 (2002). 

\bibitem{Tudor}{Tudor D. Stanescu and Philip Phillips}, arXiv:cond�mat/0104478v2 (2001).

\bibitem{Erik}{Erik G.C.P. van Loon et al.}, arXiv:1602.09129v1 [cond-mat.str-el] (2016)

\bibitem{Keldysh}{L. V. Keldysh and A. N. Kozlov}, Sov. Phys. JETP 27, 521 (1967). 

\bibitem{Plakida}{N.M.Plakida}, Phys.Rev.Lett. V.92 P.256401 (2011).

\bibitem{Zenker}{B.Zenker}, arXiv:1409.2230v1 [cond-mat.str-el], (2014)

\bibitem{Mazziotti}{LeeAnn M. Sager, Shiva Safaei, and David A. Mazziotti}, arXiv:2002.08445v1 [cond-mat.str-el] (2020)

\bibitem{Kotliar}{G.Kotliar and A.E. Ruckenstein}, Phys. Rev. Lett. 57, 1362, (1986).

\bibitem{Brech}{M. Brech, J. Voit and H. B\"utner}, Europhys. Lett., 12 (4)) 289 (1990).











\bibitem{chaschin2011_2} { N.I.Chashchin},  Phys. Met. Metallogr. {\bf 111}, 329 (2011)

\bibitem{chaschin2012_3} { N.I.Chashchin},  Phys. Met. Metallogr. {\bf 112}, 533  (2012) 

\bibitem{chaschin2016_4} { N.I.Chashchin},  Phys. Met. Metallogr. {\bf 117}, 663 (2016) 

\bibitem{chaschin2011_1} { N.I.Chashchin},  Phys. Met. Metallogr. {\bf 111},221 (2011) 

\bibitem{izyumov_Manc2005}{Yu.A. Izyumov, N.I. Chashchin N.I., D.S. Alexeev, and F. Mancini},  Eur.Phys.J.B {\bf 45}, 69 (2005)

\bibitem{chaschin2017} {N.I. Chashchin}, arXiv:707.01798v [cond-mat.str-el], (2017)

\bibitem{chaschin2020} {N.I. Chashchin}, arXiv:2011.14657 [cond-mat.str-el], (2020)

\bibitem{Meixner}{M.Meixner et al.}, Xiv:2310.17302v1 [cond-mat.str-el] (2023)

\bibitem{Pan} {Xiao-Yin Pan, Jing-Rong Wang, and Guo-Zhu Liu}, arXiv:1807.00452v2 [cond-mat.str-el], (2018)


\end{thebibliography}
\end{document}